\documentclass{ws-mpla}
\usepackage[super,compress]{cite}
\usepackage[breaklinks]{hyperref}  
\hypersetup{colorlinks,urlcolor=black,citecolor=black,linkcolor=black,filecolor=black}
\usepackage{braket}
\usepackage{graphicx} \usepackage{graphics}
\usepackage{url}
\usepackage{hyperref}\hypersetup{colorlinks=true}

\newcommand{\dd}{\mathrm{d}}
\newcommand{\ex}{\mathrm{e}}

\DeclareFontFamily{U}{mathx}{\hyphenchar\font45}
\DeclareFontShape{U}{mathx}{m}{n}{<-> mathx10}{}
\DeclareSymbolFont{mathx}{U}{mathx}{m}{n}
\DeclareMathAccent{\widebar}{0}{mathx}{"73}

\newcommand{\U}{\mathcal{U}}

\newcommand{\HA}{\mathcal{H}}

\newcommand{\GN}{G_{_\mathrm{N}}}

\newcommand{\wh}{\widetilde{\Delta\eta}}
\newcommand{\Ka}{\mathcal{K}}

\newcommand{\SV}[1]{#1}

\begin{document}

\markboth{P. Peter \& S. Vitenti}
{A simple bouncing quantum cosmological model}

\catchline{}{}{}{}{}

\title{THE SIMPLEST POSSIBLE BOUNCING QUANTUM COSMOLOGICAL MODEL}

\author{\footnotesize PATRICK PETER}

\address{Institut d'Astrophysique de Paris (${\cal G}\mathbb{R}\varepsilon\mathbb{C}{\cal O}$),
UMR 7095 CNRS,
Sorbonne Universités, UPMC Univ. Paris 06, Institut Lagrange de Paris,
98 bis boulevard Arago, 75014 Paris, France.\\
}

\author{SANDRO D.~P.~VITENTI}

\address{Institut d'Astrophysique de Paris (${\cal G}\mathbb{R}\varepsilon\mathbb{C}{\cal O}$),
UMR 7095 CNRS,
Sorbonne Universités, UPMC Univ. Paris 06, Institut Lagrange de Paris,
98 bis boulevard Arago, 75014 Paris, France.\\
}

\maketitle


\begin{abstract}
We present and expand the simplest possible quantum cosmological
bouncing model already discussed in previous works: the trajectory
formulation of quantum mechanics applied to cosmology (through the
Wheeler-De Witt equation) in the FLRW minisuperspace without spatial
curvature. The initial conditions that were previously assumed were such
that the wave function would not change its functional form but instead
provide a dynamics to its parameters. Here, we consider a more general
situation, in practice consisting of modified Gaussian wave functions,
aiming at obtaining a non singular bounce from a contracting phase.
Whereas previous works consistently obtain very symmetric bounces, we
find that it is possible to produce highly non symmetric solutions, and
even cases for which multiple bounces naturally occur. We also introduce
a means of treating the shear in this category of models by quantizing
in the Bianchi I minisuperpace.

\end{abstract}

\ccode{PACS Nos.: 98.80.Es, 98.80.-k, 98.80.Jk}

\section{Introduction}

Although recent accumulation of data\cite{Aghanim:2015xee} seem to favor
the standard\cite{Peter2009} inflationary paradigm\cite{Martin2014} with
primordial perturbations originating from quantum vacuum
fluctuations\cite{Mukhanov1992}, a few riddles still need solutions,
among which the presence of a primordial singularity\cite{Borde2003}. It
is often hoped, and shown explicitly in specific models, that quantum
gravitational effects should take care of this problem, and indeed, many
proposals have been made in this
direction\cite{Gasperini:1992em,Brandenberger:2008nx,Khoury:2001wf,Fanuel:2012ij},
although not always successfully\cite{Kallosh:2001ai,Martin2002}.

Treating the Universe itself as a quantum object immediately raises the
question of both the meaning of the wave function $\Psi$ and the issue
of measurement. Indeed, whereas a tabletop experiment leads to a simple
interpretation of $|\Psi|^2$ as the probability to measure whatever it
is one wants to measure, when one is dealing with the wave function of
the Universe, such an interpretation becomes instantly slightly more
obscure. Moreover, what is it exactly that one measures in quantum
cosmology that would collapse the wave function on the eigenvalue of?
Such questions have been the subject of long debates, including
possibilities to modify quantum mechanics itself by inclusion of
non linear and stochastic terms in a new version of the Schr\"odinger
equation.\cite{Martin2012,Leon:2015eza}

\SV{Similarly using a quantum cosmological framework in which the usual
quantum mechanics is modified}, in Ref.~\refcite{AcaciodeBarros1998}, a
plausible solution to all questions and issues above was shown to
possibly stem from assuming an actual (quantum)
motion\cite{deBroglie:1927,Bohm:1951xw,Holland1993} of the scale factor
$a$ for quantum cosmology expressed in the
Friedmann-Lema\^{\i}tre-Robertson-Walker (FLRW) minisuperspace:
quantizing a barotropic fluid along with the gravitational field, one
obtains a well-defined time variable\footnote{This is achieved through a
formalism introduced by Schutz\cite{Schutz:1970my,Schutz:1971ac} in
which the velocity potential can be expressed in terms of a field $T$
whose canonical conjugate $p_T$ appears linearly in the Hamiltonian,
thus providing the basis for a time-dependent Schr\"odinger equation.}
$T$ entitling to rewrite the Wheeler-De Witt equation in a Schr\"odinger
form to which a de Broglie-Bohm (dBB) trajectory $a(T)$ can then
assigned. With either dust or radiation, and with arbitrary spatial
curvature, one sees that for a given initial (essentially gaussian) wave
function, the resulting trajectory never goes through $a=0$, and hence
that no singularity is ever reached\SV{: a bounce instead takes
place.\cite{Battefeld:2014uga}}

In what follows, we present an extension of this result\footnote{Another
extension was presented in Ref.~\refcite{PintoNeto:2009gv} with a different
goal related with the initial probability of a large Universe.} aiming at
generalizing the initial condition wave function $\Psi_0$ given at an
arbitrary time: in the previous work\cite{AcaciodeBarros1998}, the
initial wave functions that were used had been chosen in such a way
that the only effect of the time evolution was to give a
time dependency on the parameters without modifying their
functional forms. For example, for a Gaussian initial condition, the
time evolution just modifies the variance, adding a complex value to it.
Here we assume more general initial wave packets for which this simple
behavior does not hold.

To simplify matters, we restrict attention to flat spatial curvature
(although in full generality this component should be included), but
allow for the presence of shear and thus start with the Bianchi I
Hamiltonian\cite{Bergeron:2015jpa}. Splitting the wave function into
a function of the scale factor only and a shear component eigenmode, we
then reduce our system to the simplest
possible case equivalent to the free particle, which we then endow with
a complex initial wave function. Building the Hilbert space, we
calculate the propagator and subsequently evolve this initial wave
function to derive the necessary equation of motion of the scale factor.
We then solve the dBB trajectory numerically to show that non symmetric
bouncing solutions appear generic in this model. We conclude by
discussing the generality of our result and providing the future
investigation directions.

\section{A Very Simple Hamiltonian}

Our starting point will be the Hamiltonian, discussed in
Ref.~\refcite{Bergeron:2015jpa}, obtained to study the evolution of the
anisotropic and compact locally flat spacetime with line element
\begin{equation}
\dd s^2 = -N^2(t) \dd t^2 + \sum_{i=1}^3 a_i^2(t) \left( \dd x^i\right)^2,
\label{metric}
\end{equation}
and matter consisting in a radiation fluid with equation of state
$w\equiv p/\rho = \frac13$. \SV{In full generality, one could choose
any value for $w$, leaving, as we will see below, the time parameter
unfixed, but here we assume that we are considering extremely early
stages of the Universe during which, if a fluid description is to make
any sense at all (and that it questionable indeed), it will have to be
a radiation fluid; setting $w=\frac13$ is therefore, in our view, not
restricting to a particular case but merely a physically motivated
statement.}

In the framework above, one sees that the relevant Hamiltonian provided
by the gravitational part of the action (the fluid part is discussed in
Sec.~\ref{sec:time}), assuming the conformal time
choice $N\to a$ (and therefore $t\to \eta$ in what follows), reads
\begin{equation}
H = \frac{\Pi_a^2}{24} - \frac{p_-^2 + p_+^2}{24 a^2},
\label{Hgrav}
\end{equation}
in terms of the canonical variables $a\equiv \left( a_1 a_2
a_3\right)^\frac13$ (average scale factor) and its associated momentum
$\Pi_a$, and the momenta $p_-$ and $p_+$, respectively conjugate of
\SV{the shear-inducing variables}
$\beta_- \equiv \frac{1}{2\sqrt{3}}\ln \left( a_1/a_2\right)$ and $\beta_+
\equiv \frac16\ln\left( a_1 a_2/a_3^2\right)$. Quantization is achieved
by promoting these variables to operators in a Hilbert space (defined
below) satisfying the usual canonical commutation relations $[ \hat
a,\hat \Pi_a] =[ \hat \beta_\pm,\hat p_\pm] = i$ (we work in
geometrical units in which $\hbar=c=\GN=1$).

\subsection{Bianchi I}

In order to move ahead, we first rescale the variables in such a way
that they retain their commutation relations and thus perform the
replacements $a\to \hat{a}/(2\sqrt{6})$ and $\Pi_a \to 2\sqrt{6} \hat{\Pi}_a$,
leading to
\begin{equation}
\hat H = \hat \Pi_a^2 - \left( \hat p_-^2 +\hat p_+^2 \right)\hat a^{-2},
\label{Haml1}
\end{equation}
which has to be compared with the usual FLRW case with no restriction on
the spatial curvature $\mathcal{K}$ for which the last term is
proportional to $\mathcal{K}\hat a^2$ (the difference, quartic in
the scale factor, is the same that holds between the curvature and shear
terms appearing in the classical Friedman equation in which the
curvature terms is merely $\Ka/a^2$ while the shear energy density is
$\propto a^{-6}$; see Ref.~\refcite{Battefeld:2014uga} and references
therein).

We now work in a mixed representation for the wave function in which the
operators $\hat a$ and $\hat p_\pm$ are multiplication operators, their
action on the wave function yielding mere multiplication by $c-$numbers,
i.e., $\hat a \Psi = a \Psi$ and $\hat p_\pm \Psi = p_\pm \Psi$, so that
the conjugate operators read $\hat \Pi_a = -i\partial/\partial a$ and
$\hat \beta_\pm = i\partial/\partial p_\pm$; this is a position
representation for the scale factor part and the momentum representation
for the shear. Since the average scale factor $a$ is, by construction, a
positive quantity, contrary to $p_\pm$ which are {\sl a priori} arbitrary
real numbers, the Hilbert space $\mathbb{H}$ is contained in the set of
square integrable functions of $\mathbb{R}^+$ and $\mathbb{R}^2$,
namely
\begin{equation}
\mathbb{H} \subset \Set{ f\left( a, p_+, p_-\right)\in \mathbb{C} | \int_0^\infty \dd a
\int_{-\infty}^\infty \dd p_+ \int_{-\infty}^\infty \dd p_- |f\left( a, p_+, p_-\right)|^2
<\infty}.
\label{Hilbert}
\end{equation}
The eigenvalue equation $\hat H \Psi = \ell^{2} \Psi$ (we note
$\ell^{2}$ the eigenvalue of the Hamiltonian operator for later
convenience) then transforms into
\begin{equation}
-\frac{\partial^2 \U^{(k)}_\ell}{\partial a^2} -\frac{k^2}{4 a^2} \U^{(k)}_\ell
= \ell^2 \U^{(k)}_\ell,
\label{eigena}
\end{equation}
where we set $k^2 \equiv 4(p_+^2 + p_-^2)$ and we have expanded the
total wave function in terms of the eigenstates $\chi(\beta_\pm)$ of
$\hat \beta_\pm$, i.e. $\chi\propto \exp[i(p_+\beta_+ + p_-\beta_-)]$, as
\begin{equation}
\Psi\left( a,p_\pm\right) = \int_0^\infty \dd\ell \int_{-\infty}^\infty \dd\beta_+
\int_{-\infty}^\infty \dd\beta_- \tilde \Psi\left(\ell,\beta_\pm\right)
\ex^{i[\beta_+ p_+ +\beta_- p_-]} \U^{(k)}_\ell(a).
\label{PsiExp}
\end{equation}
Our Hilbert space will then be completed by obtaining the solutions for
the energy eigenmodes $\U^{(k)}_\ell(a)$.

\subsection{Hilbert space and boundary conditions}

We now implement the requirement that the Hamiltonian should be
self-adjoint, namely that
\begin{equation}
\int\dd a\,\dd^2p \, \left(H\Psi\right)^* \Psi = \int\dd a\,\dd^2p \,  
\Psi^* \left(H\Psi\right),
\label{HHp}
\end{equation}
with the limits of integration defined in \eqref{Hilbert}. As already
mentioned above, the set of normalizable functions is too large and the
actual Hilbert space $\mathbb{H}$ must be restricted to those
normalizable functions such that the condition above is satisfied. We
note that the condition \eqref{HHp} is automatically satisfied if the
Hamiltonian eigenvectors form an orthonormal basis, i.e.
\begin{equation}
\int_0^\infty \dd a \, \U^{(k)*}_\ell (a) \U^{(k)}_{\ell'} (a) = \delta(\ell-\ell')
\label{Ull}
\end{equation}
and
\begin{equation}
\int_0^\infty \dd\ell \int_{-\infty}^\infty \dd\beta_+
\int_{-\infty}^\infty \dd\beta_-| \tilde \Psi\left(\ell,\beta_\pm\right) |^2 \ell^2
< \infty.
\label{noninft}
\end{equation}
We therefore define our Hilbert space $\mathbb{H}$ to be the set of
square integrable functions of the form \eqref{PsiExp} where the basis
is orthonormal in the sense of~\eqref{Ull} and the functions
$\tilde\Psi\left(\ell,\beta_\pm\right)$ satisfy Eq.~\eqref{noninft}.

Setting $\nu = \frac12\sqrt{1-k^2}$, we obtain the general solution for
the energy eigenmodes
\begin{equation}
\U^{(k)}_{\ell} (a) = c_+ \sqrt{a\ell} J_\nu (a \ell) +
c_- \sqrt{a\ell} J_{-\nu} (a \ell),
\label{ElJnu}
\end{equation}
in terms of the Bessel functions $J_\nu$, with $c_\pm$ complex numbers
satisfying $|c_+|^2 + |c_-|^2 = 1$. Imposing the condition \eqref{Ull}
then demands that either $c_+$ or $c_-$ vanishes. \SV{Apart from an
irrelevant phase factor, one can therefore, without lack of generality, 
set $c_+=1$ and $c_-=0$, or the opposite $c_+=0$ and $c_-=1$; given
the initial wave function choice, this will lead to impose different
boundary conditions.}

\section{Time Evolution}
\label{sec:time}

As alluded to in the introduction, the existence of a fluid implies that of a
preferred time slicing, and hence of a natural time parameter $t$ for
the Schr\"odinger equation. Following Schutz
formalism\cite{Schutz:1970my,Schutz:1971ac}, the velocity potential
of the fluid allows to define a field whose properties are similar to that
expected for a time variable. One then needs to fix the coordinate
time variable appearing in \eqref{metric} through a clever choice of
the lapse function $N(t)$ adapted to the fluid. The most appropriate choice 
in \eqref{metric} happens to depend on the equation of state of
the fluid, and for a radiation dominated universe, one can set $N=a$, so
the relevant time is the usual conformal time $\eta$, in terms of which
the fluid canonical momentum reads $\hat P_\mathrm{fluid}
=-i\partial_\eta$. This momentum enters the Hamiltonian linearly,
so that the Wheeler-De Witt equation reads like a simple Sch\"odinger equation,
resulting in an evolution operator $U(\eta,\eta_0)$ satisfying
\begin{equation}
i\frac{\partial U}{\partial\eta} = \hat H U \ \ \ \ \Longrightarrow
\ \ \ \ \ U(\eta,\eta_0) = \ex^{ -i\left( \eta-\eta_0\right)\hat H},
\label{propdef}
\end{equation}
with $\eta_0$ an arbitrary initial time. Anticipating on the dBB
trajectory, we note that, contrary to previous works, we do not intend
here to impose initial condition at the expected bounce time at which
$\dd a/\dd \eta =0$, assuming such a time to exist, but we will instead
demand arbitrary initial condition in a contracting stage $\dd a/\dd
\eta <0$ and actually evolve the universe until it bounces \SV{(again
anticipating that it will do so)}.

\subsection{Propagator}

Our ultimate goal is, starting from an arbitrary initial wave function
$\Psi_0 (a) = \Psi(a;\eta_0)$, to follow its time development
$\Psi(a;\eta)$, whose phase $S(a;\eta)$ provides the dBB trajectory
through
\begin{equation}
\frac{\dd a}{\dd\eta} = \frac{\partial S}{\partial a} = \frac{i}{2|\Psi|^2} \left(
\Psi\frac{\partial\Psi^*}{\partial a} - \frac{\partial\Psi}{\partial a}\Psi^* \right),
\label{dadeta}
\end{equation}
and to figure under which conditions this time evolution, starting from
a contracting scale factor, leads to a regular bouncing solution. We
shall deal with the initial condition in the forthcoming section, and
for now on concentrate on the propagator $G(a,p_\pm,a_0,p^0_\pm)\equiv
\langle a, p_\pm|U|a_0,p^0_\pm\rangle$ evolving the state
$|a_0,p^0_\pm\rangle$ at time $\eta_0$ into $|a,p_\pm\rangle$ at time
$\eta$.

In order to obtain this propagator, we first express the time evolution
operator on the Fourier modes as
\begin{equation}
U = \int\dd^2\beta\! \int_0^\infty \!\dd\ell\, U(\eta,\eta_0)
|\ell,\beta_\pm\rangle \langle \ell,\beta_\pm| =
\int\dd^2\beta\!\int_0^\infty \!\dd\ell \, \ex^{-i\ell^2 \Delta\eta}
|\ell,\beta_\pm\rangle \langle \ell,\beta_\pm|,
\label{Ubase}
\end{equation}
with $\Delta\eta \equiv \eta-\eta_0$. This leads to
\begin{equation}
G(a,p_\pm,a_0,p^0_\pm) = \delta^{(2)} (p_\pm-p'_\pm) \int_0^\infty \!\dd\ell \,
\ex^{-i\ell^2 \Delta\eta} \U_\ell^{(k)}(a) \U_\ell^{(k)*}(a'),
\label{Gaa}
\end{equation}
where we have used $\langle a,p_\pm|\ell,\beta_\pm\rangle =
\U_\ell^{(k)}(a) \exp[-i(p_+\beta_+ + p_- \beta_-)]/(2\pi)$. As
expected, we see that if one starts with an eigenstate of $\hat p_\pm$,
the subsequent time evolution remains on this state. From now on, we
shall therefore discard the shear contribution insofar as it does not
contribute to the equation of motion and merely assume the value
of the shear to be fixed; we therefore simply denote the
propagator by $G(a,a_0;\eta)$ in what follows.

The propagator \eqref{Gaa} needs be regularized, which is done by
replacing $\Delta\eta$ by $\wh = \Delta\eta (1+i\epsilon)$, with
$\epsilon\ll 1$. Then, using Eq.~(10.22.67) of Ref.~\refcite{Olver:2010:NHMF}, one
finds
\begin{equation}
G(a,a_0;\eta) = -\frac{i \sqrt{a a_0}}{2\wh} \ex^{\frac{i}{4}(a^2+a_0^2)/\wh -i\alpha\pi/2}
J_\nu\left(\frac{a a_0}{2\wh}\right),
\label{Gaaeta}
\end{equation}
thanks to which we are now in a position to derive the wave function
at an arbitrary time.

\begin{figure}[t]
\centerline{\includegraphics[width=15cm]{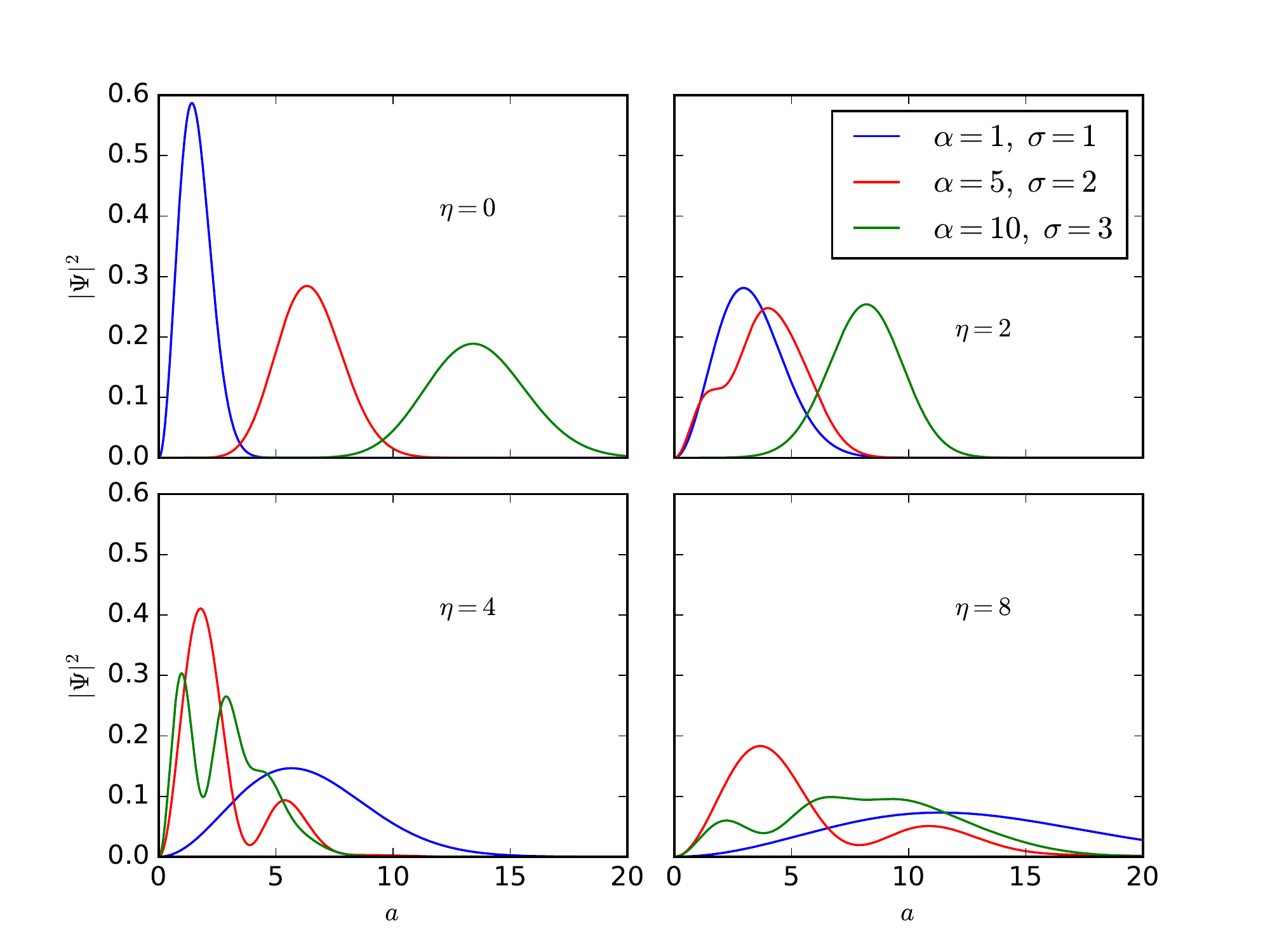}} 
\caption{Initial wave function and its time evolution for various
values of the localization $\alpha$ and spread $\sigma$. We used
$\HA_\mathrm{ini} = -0.1$ for all wave functions and $\nu = \frac12$,
sticking to the FLRW case to evolve the wave functions. As expected, the
wave form changes dramatically during the time evolution, as in particular
the mode, depending on its initial localization and spread, eventually
reaches the potential wall at $a=0$ at which point the wave function
bounces off. Oscillations then appear, due to the reflection on the
wall, inducing a significant change in the subsequent time evolution; 
these changes are responsible, as we show below, for the completely
new behavior of the scale factor exhibited on Fig.~\ref{figaeta}}
\protect\label{figPsi02}
\end{figure}

\subsection{Wave function}

Having set the general framework for Bianchi I, we now simplify once
more the model to restrict attention to the flat FLRW case, which
represents a subset of Bianchi I; the complete analysis of the Bianchi I
case will be presented elsewhere in a future work\cite{PPSVPrep}. As we
discuss below, the wave function proposed in
Ref.~\refcite{AcaciodeBarros1998} can be generalized to yield new
bouncing solutions. We shall accordingly consider an
initial wave function at time $\eta_0$ which we demand to be localized
around a certain initial value $a_\mathrm{ini}$ of the scale factor and
having a complex phase in order to account for a possible initial
velocity. Generalizing Ref.~\refcite{AcaciodeBarros1998}, we set the
normalized form
\begin{equation}
\Psi_0(a) = \langle a,p_\pm| \Psi_0\rangle = \frac{2^{(1-2\alpha)/4}\ a^\alpha }{\sigma^{\alpha+1/2}\sqrt{\Gamma\left(\alpha+\frac12 \right)}}
\exp \left[-\frac12 a^2 \left(
\frac{1}{2\sigma^2} - i \HA_\mathrm{ini} \right)\right],
\label{Psi0}
\end{equation}
where $\alpha$, $\sigma$ and $\gamma$ are real but otherwise arbitrary
parameters. The wave function \eqref{Psi0} peaks at $a_\mathrm{ini} =
\sqrt{2\alpha} \sigma$ (maximum of $|\Psi_0|^2$), has a mean value at
$\bar a = \langle a \rangle = \sqrt{2}\sigma \Gamma\left(1+\alpha\right)
/\Gamma\left(\frac12+\alpha\right)$ and variance $\langle \left(a-\bar
a\right)^2\rangle =
\sigma^2\left[1+2\alpha-2\Gamma^2\left(1+\alpha\right)
/\Gamma^2\left(\frac12+\alpha\right)\right]$. As shown on
Fig.~\ref{figPsi02}, the peak localization is essentially given by
$\alpha$ while the spread mostly stems from the value of $\sigma$.
Finally, applying \eqref{dadeta} to the initial $\Psi_0$ \eqref{Psi0},
one finds that the coefficient $\HA_\mathrm{ini}$ actually
represents the initial value of the conformal Hubble parameter,
hence the name of the parameter.

Equipped with the initial wave function \eqref{Psi0} and the propagator
\eqref{Gaaeta}, we are now in position to derive the equation of motion
for the dBB scale factor. The wave function at time $\eta$ is 
\begin{equation}
\Psi(a;\eta) = \langle a|U(\eta,\eta_0)|\Psi_0\rangle \propto
\int_0^\infty \!\dd b\, \sqrt{ab} \ex^{\frac{i}{4} (a^2+b^2)/\wh} J_\nu\!\left(\frac{ab}{2\wh}\right) b^\alpha \ex^{-\frac12 b^2 \left(
\frac{1}{2\sigma^2} - i \HA_\mathrm{ini} \right)},
\label{Psieta}
\end{equation}
which, although it happens to be explicitly integrable in terms of
hypergeometric functions, is not particularly illuminating.
Figure~\ref{figPsi02} also shows the typical time evolution
\eqref{Psieta} for the modulus of the wave function. This evolution
clearly differs from that of Ref.~\refcite{AcaciodeBarros1998} where it
would hold its analytical shape at all times with time-dependent parameters. Here,
one sees that the boundary condition at $a=0$ acts as an infinite
potential wall such that, when the wave function evolves towards it, it
bounces off, thereby inducing subsequent oscillations that can change
dramatically the dBB trajectories, as we now discuss.

\section{Results: from Bianchi I to FLRW}

Let us move on to the results, and for the purpose of exemplifying,
concentrate on the shearless limit whereby $\nu\to\frac12$. The full
analysis of the Bianchi I case will be provided
elsewhere\cite{PPSVPrep}, and for the purpose of this work, we will make
contact with Ref.~\refcite{AcaciodeBarros1998} by going to the vanishing
shear limit for which $k=0$, and hence $\nu=\frac12$, so the basis
simplifies to mere sines and cosines. In Ref.~\refcite{AcaciodeBarros1998},
the requirement that $\hat H$ be self-adjoint was shown to lead to the
condition
\begin{equation}
\lim_{a\to 0} \frac{\dd \Psi(a)}{\dd a} = \lim_{a\to 0} \alpha_\mathrm{BC} \Psi(a),
\end{equation}
with $\alpha_\mathrm{BC} \in (-\infty,\infty)$. The cases studied then
correspond respectively to $\alpha_\mathrm{BC} \to 0 \Longrightarrow
(c_-=1, c_+=0)$ and $\alpha_\mathrm{BC}\to \infty \Longrightarrow
(c_-=0, c_+=1)$, as already mentioned. The Hamiltonian reduces to
that of a free particle on the half-line with the point $a=0$
equivalent to an infinite potential wall. In our final result
\eqref{Psieta}, the Bessel function then reduces to a hyperbolic sine of
its argument, so that most calculations are analytically tractable.
We shall not go along this direction here\cite{PPSVPrep} and instead
concentrate on a numerical evaluation of the dBB trajectories, comparing
those with previous calculations.

\begin{figure}[t]
\centerline{ 
\includegraphics[height=5cm]{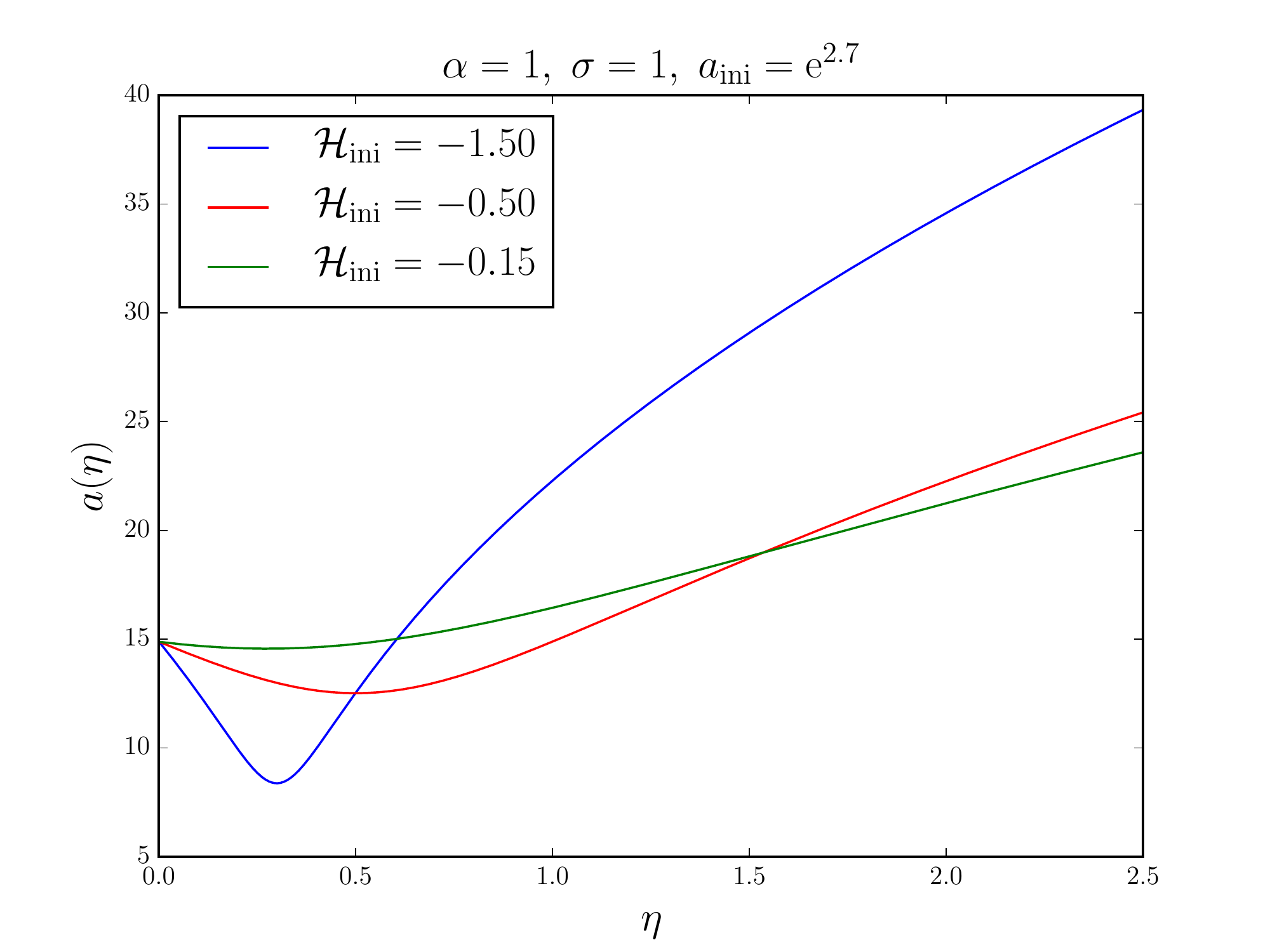}
\includegraphics[height=5cm]{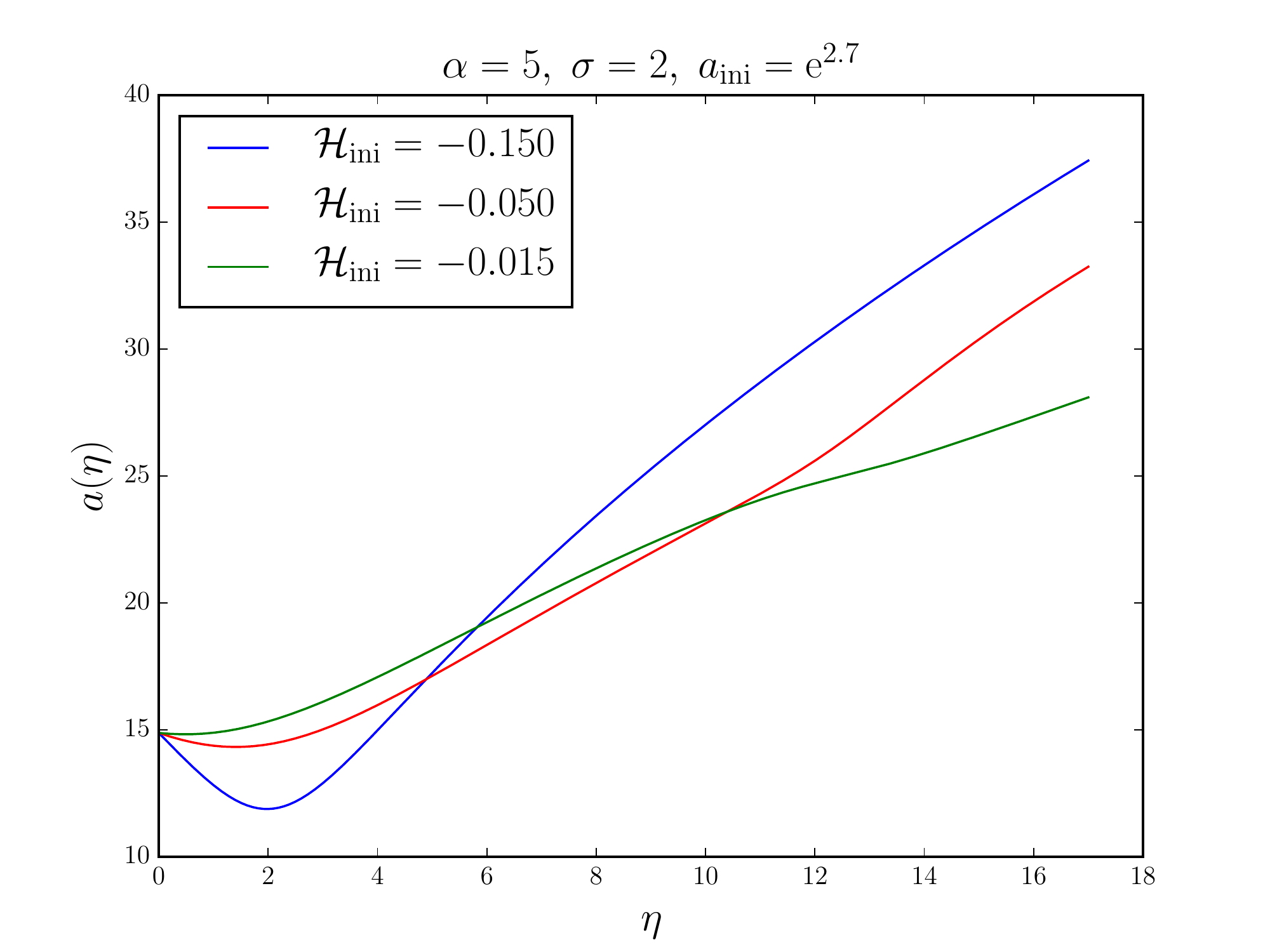} }
\centerline{
\includegraphics[height=5cm]{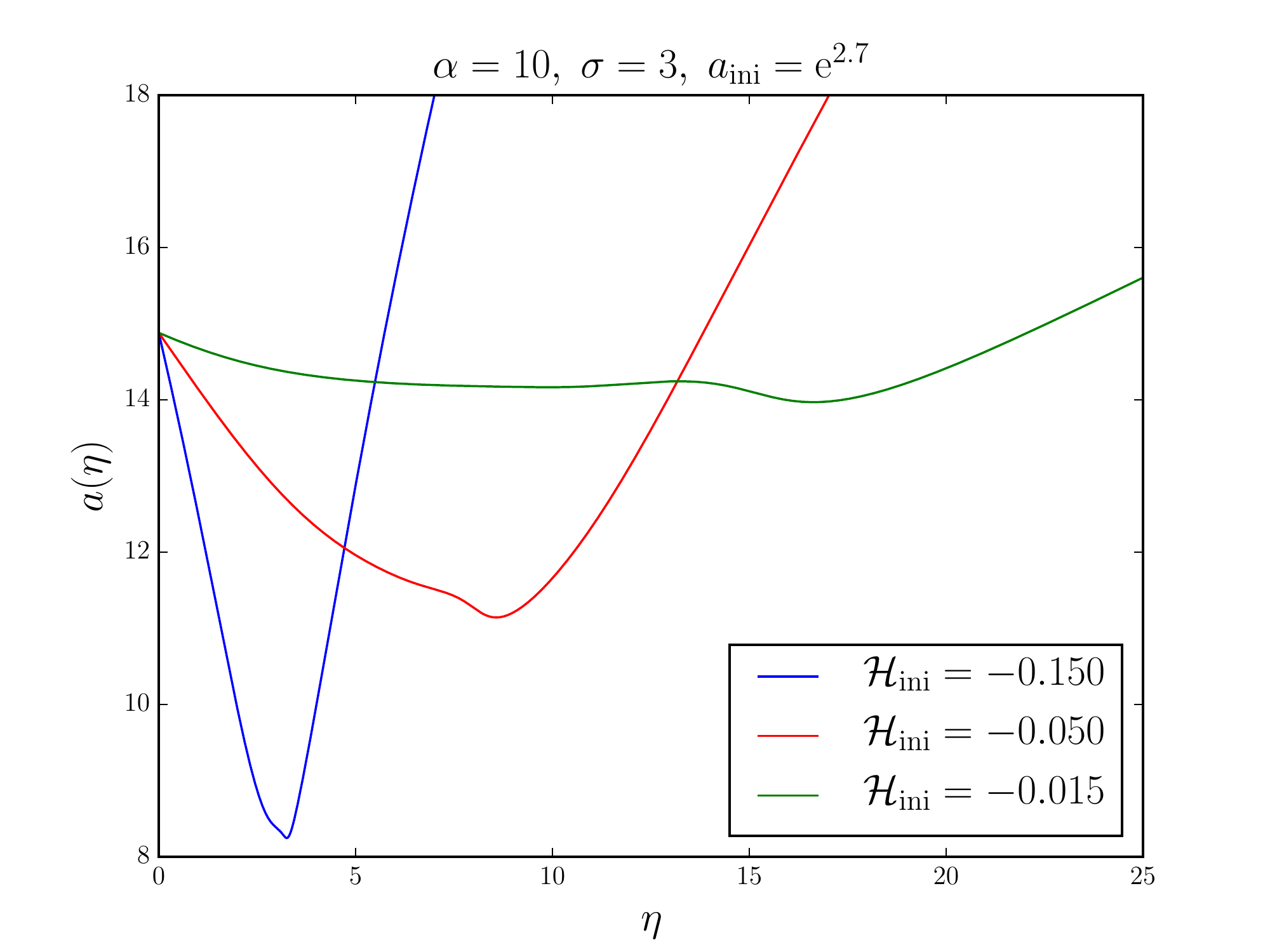}
\includegraphics[height=5cm]{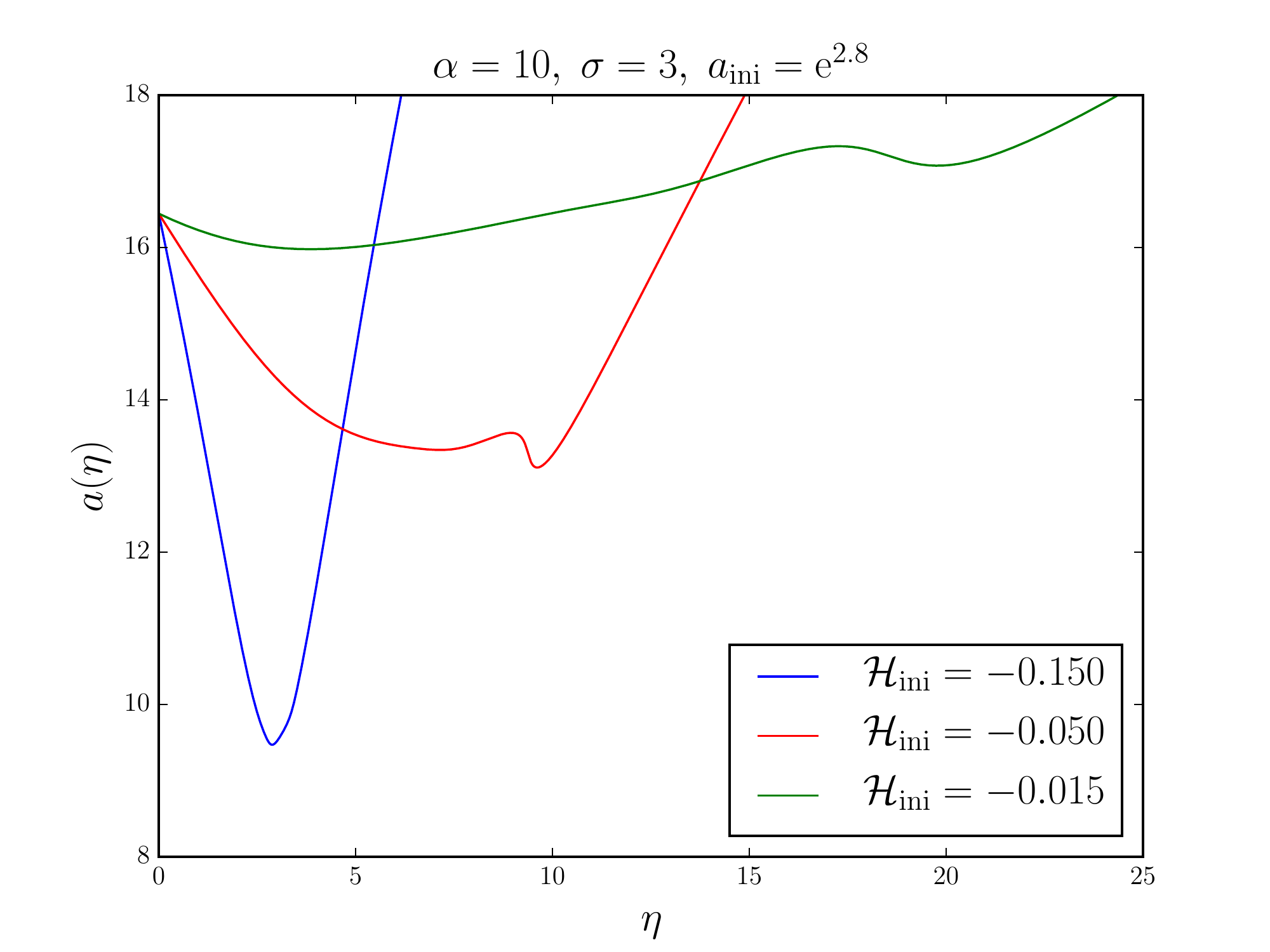}
} 
\caption{Illustration, for the shearless case $\nu=\frac12$, of
typical bouncing models obtained as solutions of the dBB
equation of motion for the scale factor. The curves are labeled
with the initial condition assumed for $\HA_\mathrm{ini}$, and
the solutions are obtained for values of $\alpha$ and $\sigma$
as in Fig.~\ref{figPsi02}. One clearly sees that not only are
such bouncing solutions not necessarily symmetric in time, but
also that they can produce many bounces and not just one, even
in the oversimplified situation of the flat shearless FLRW.}
\protect\label{figaeta}
\end{figure}

The pioneering solutions obtained in Ref.~\refcite{AcaciodeBarros1998}
for the shearless case with no spatial curvature depend on two parameters,
denoted $\eta_*$ and $a_\mathrm{B}$, and read
\begin{equation}
a(\eta) = a_\mathrm{B} \sqrt{1+\left(\frac{\eta}{\eta_*}\right)^2};
\label{sol0}
\end{equation}
these parameters can be understood respectively as the typical bounce duration
($\eta_*$) and the minimum value of the scale factor ($a_\mathrm{B}$).
This family of solution has very simple properties: first, as expected
from a quantum gravity framework, they solve the singularity problem
(what would be the point to have a quantum gravity model with
singularities?) in the sense that the contracting phase never reaches the
singular point $a=0$ since the minimum scale factor $a_\mathrm{B}>0$
unless one sets $a_\mathrm{B}=0$ in the first place, in which case the
solution is singular at all times and thus lacks any physical relevance.

A second conclusion that can be drawn from \eqref{sol0} is that there is
one and only one bounce taking place, with a regular decrease of the
scale factor followed by a simple bounce and a subsequent regular
increase. In a sense, this is the simplest extension of the standard
cosmological model that can be thought of. Finally, and this is possibly
the most crucial point, the bounces induced by \eqref{sol0} are all
symmetric in time.

Figure~\ref{figaeta} shows various solutions assuming different values
for the initial scale factor and conformal Hubble parameter, for the
cases depicted on Fig.~\ref{figPsi02}. It is apparent that the very
simple solution of Ref.~\refcite{AcaciodeBarros1998}, is merely an
exceptional case. In the cases studied here, we find that not only is
the bounce often non symmetric in time, but also that many bounces can
naturally occur, and in a way which is quite sensitive to the initial
condition one sets on $a_\mathrm{ini}$ and $\HA_\mathrm{ini}$.

Our study opens a wide range of new studies that need now be done for a
complete understanding of such bouncing scenarios. First, the case with
non vanishing shear must be investigated in details, with particular
emphasis on the so-called BKL instability (see again
Ref.~\refcite{Battefeld:2014uga} for a thorough discussion of bounce
models and their problems) according to which a pre-existing shear can
spontaneously lead to many new Kasner-like singularities.

The second point that should be examined in details concerns the
propagation of perturbations through such a complicated bounce. In
models such as those based on dBB trajectories, it was found that
perturbations can be easily evaluated in a way reminiscent the ordinary
perturbation theory based on general relativity, but with the scale
factor replaced by that obtained along the dBB
trajectory\cite{Peter2005,Pinho2007,Peter2007}. With characteristic
solutions exhibiting such features as shown on Fig.~\ref{figaeta}, it is
clear that the potential for the perturbations will have many new
interesting consequences that we hope to clarify in a further extension
of the current work.

\section*{Acknowledgments}

We thank CNPq of Brazil and ILP (Ref. ANR-10-LABX-63) for financial support.

\bibliographystyle{ws-mpla}

\end{document}